\documentclass[11pt]{article}
\title{A bird's-eye view of nonlinear-optical processes: unification through scale invariance}
\author{Mark G. Kuzyk  \\ Department  of Physics and Astronomy, Washington State University \\ Pullman, WA  99164-2814  \\ kuz@wsu.edu}
\usepackage{graphics}
\usepackage{amsmath}
\begin{document}
\maketitle

\begin{abstract}
The Schr\"{o}dinger equation has the property that when changing the length scale by $\vec{r} \rightarrow \epsilon \vec{r}$ and the energy scale by $E \rightarrow E / \epsilon^2$, the shape of the wavefunction remains unchanged.  The same re-scaling leaves the intrinsic hyperpolarizability (as well as higher-order hyperpolarizabilities) unchanged.  As such, the intrinsic hyperpolarizability is the best quantity for comparing molecules since it re-normalizes for trivial differences that are due to molecular size and energy gap.  Similarly, the intrinsic hyperpolarizability is invariant to changes in the number of electrons.  In this paper, we review the concept of scale invariance and how it can be applied to better understand the nonlinear optical response, which can be used to develop new paradigms for it's optimization.
\end{abstract}

\section{Introduction}

A decade ago, the fundamental limits of the nonlinear-optical susceptibilities were calculated.\cite{kuzyk00.01,kuzyk00.02} These calculations were based on the sum rules, which are equations that relate transition moments and energies to each other, to simplify the sum-over-states (SOS) expressions.\cite{orr71.01}  Since the polarizability is rigourously optimized when all of the oscillator strength is concentrated in one transition, one might expect the same to be true for the nonlinear susceptibilities.  However, the sum rules show that this approach leads to inconsistencies unless at least two excited states are used.

{\em The three-level anzatz} (TLA) states that only three states contribute to the nonlinear-optical response when the hyperpolarizability is at the fundamental limit.\cite{kuzyk05.01,zhou06.01,kuzyk06.03,zhou07.02} The TLA, in conjunction with the sum rules, was used in calculating the fundamental limits.  While the three-level ansatz has not yet been rigourously proven, it is always observed to hold when the nonlinear response of a molecule is optimized.  Indeed, even Monte Carlo methods, which randomly assign transition moments and energies while enforcing the sum rules, show that the largest hyperpolarizability calculated using the dipole-free SOS expression always obeys the three-level ansatz.\cite{kuzyk08.01} Given that this approach makes no assumptions about the form of the wavefunctions suggests that the three-level ansatz may be universally true.

The fundamental limits of the nonlinear susceptibilities define absolute standards to which one can compare the efficiency of a molecule.\cite{zhou08.01}  In Section \ref{sec:scale invarience}, we show that the ratio of the nonlinear susceptibility to the fundamental limit - called the intrinsic $n^{th}$-order intrinsic hyperpolarizability, is a scale invariant quantity that does not change when the size of a molecule is changed in a way that preserves the shape of the wavefunctions.  It is also independent of the number of electrons in the system. As such, the intrinsic hyperpolarizabilities remove the effects of simple scaling, and should therefore be the property of focus in structure-property studies.

In Section \ref{sec:application}, we show how an analysis using universal diagrams of a large collection of molecules with the intrinsic hyperpolarizability as the dependent variable leads to a deeper understanding of the intrinsic properties of a material that are responsible for optimizing the nonlinear response.  This type of analysis leads to the proposal that modulation of conjugation may be one method for making better molecules that exceed the long-standing intrinsic hyperpolarizability ceiling of 0.03.

In Section \ref{paradigms}, we show that certain universal features are shared by all quantum systems whose nonlinear susceptibilities are at the fundamental limit.  For example, extensive numerical simulations suggest that an optimized molecule must have two and only two dominant excited states.  Furthermore, the ratio of energies between the two excited states, $E_{10}/E_{20}$, must be about 0.49; and, the normalized transition moment to the first excited state must be 0.709.  This suggests that underlying unifying principles may be at work, and raises new questions about why nature appears to require one and only one set of properties to optimize the nonlinear response.  From a pragmatic perspective, these universal properties provide a set of criteria for searching databases of molecular properties to identify new structures that may have an enhanced nonlinear-optical response.

\section{Scale Invariance}\label{sec:scale invarience}

The most general form of the Hamiltonian for N particles of mass $m$ is,
\begin{equation}\label{generalHamiltonian}
H = \frac {1} {2m} \sum_{k=1}^{N} \left( \vec{p}_k - \frac {e} {c} \vec{A} \left(\vec{r}_1, \dots \vec{r}_N \right) \right)^2 + V \left(\vec{r}_1, \dots \vec{r}_N; \vec{s}_1, \dots \vec{s}_N ; \vec{L}_1, \dots \vec{L}_N\right),
\end{equation}
where $\vec{p}_k$ is the momentum of particle $k$, $e$ is the particle's charge, $c$ the speed of light, $\vec{A}$ the vector potential, $V$ the scalar potential; and, where $\vec{r}_k$, $\vec{L}_k$ and $\vec{s}_k$ are the positions, orbital angular momenta and spins of the $k^{th}$ particle.  The potentials and momenta of the particles are functions of the positions.

The Hamiltonian given by Equation \ref{generalHamiltonian} describes any N-electron system, such as a molecule or charges in a multiple quantum well.  For example, the potential, $V$, can contain coulomb repulsion terms, such as $-e^2/\left|\vec{r}_1 - \vec{r}_2 \right|$, spin interactions of the form $v(\left| \vec{r}_1 - \vec{r}_2 \right|) \vec{s}_1 \cdot \vec{s}_2$, spin-orbit coupling, and external electric fields; while the vector potential can describe interactions with external electric and magnetic fields.

The N-electron Schr\"{o}dinger equation is of the form,
\begin{equation}\label{schrodinger}
H \psi(\vec{r}_1, \dots \vec{r}_N) = E \psi(\vec{r}_1, \dots \vec{r}_N).
\end{equation}
When Equation \ref{schrodinger} is transformed by $\vec{r}_k \rightarrow \epsilon \vec{r}_k$ we get,
\begin{eqnarray}\label{schrodingerRe-scaled}
&& \left[ \frac {1} {2m} \sum_k^{N} \left( \vec{p}_k - \frac {e} {c} \epsilon \vec{A} \left(\epsilon \vec{r}_1, \dots \right) \right)^2 \right. \nonumber \\ & + & \left. \epsilon^2 V \left(\epsilon\vec{r}_1, \dots; \vec{s}_1, \dots; \vec{L}_1, \dots \right) \right] \psi(\epsilon \vec{r}_1, \dots ) = \epsilon^2 E \psi(\epsilon \vec{r}_1, \dots ),
\end{eqnarray}
where we have used $\vec{p}_k \rightarrow \vec{p}_k / \epsilon^2$ since in the position representation, $\nabla_k^2 \rightarrow \nabla_k^2 / \epsilon^2 $.  Clearly, $\psi(\epsilon \vec{r}_1, \dots)$ - which has the same shape as $\psi(\vec{r}_1 , \dots)$ aside from being spatially compressed by a factor $1/\epsilon$ - is a solution of the Schr\"{o}dinger equation with $E \rightarrow E \epsilon^2$, $V(\vec{r}) \rightarrow V(\epsilon \vec{r}_1, \dots) \epsilon^2$, and $\vec{A} \left( \vec{r}_1, \dots \right)  \rightarrow \epsilon \vec{A} \left(\epsilon \vec{r}_1, \dots \right) $.  Thus, compressing the potentials spatially by a factor of $1/\epsilon$, re-scaling the energy by $\epsilon^2$, and re-scaling the vector potential by $\epsilon$ leaves the shape of the wavefunctions unchanged.  By re-scaling, we specifically mean the transformation of Equation~\ref{schrodinger} into Equation~\ref{schrodingerRe-scaled} with the associated re-scaling of the energies and vector potential.

Upon re-scaling, the position and energy product $\vec{r} \cdot \vec{r} E$ remains invariant, or
\begin{equation}\label{InvarientProduct}
\vec{r}\cdot \vec{r} E = constant ,
\end{equation}
where $\vec{r}$ is defined as follows.  The dipole moment, $\vec{\mu}$ is defined as,
\begin{equation}\label{dipole}
\vec{\mu} = -e \sum_i^N \vec{r}_i,
\end{equation}
where $\vec{r}_i$ is the position of the $i^{th}$ electron, and $-e$ the electron charge.  Thus, the position operator is defined by,
\begin{equation}\label{position}
\vec{r} = - \frac {\vec{\mu}} {e} .
\end{equation}

The sum rules are calculated from the fact that the commutator of the Hamiltonian given by Equation \ref{generalHamiltonian} with any component of the position operator yields (see appendix),
\begin{equation}\label{commutator}
\left[x,\left[ x,H \right] \right] = \frac {\hbar^2} {m} .
\end{equation}
The resulting sum rules follow by applying closure ($\sum_{n} \left| n \right> \left< n \right| = 1$) to Equation \ref{commutator}, yielding,\cite{kuzyk01.01}
\begin{equation}\label{therule}
\frac {2m} {\hbar^2} \sum_{n}  x_{mn} x_{np} \left(  E_n - \frac {1} {2} \left( E_p + E_m \right) \right) = N \delta_{m,p},
\end{equation}
where $x_{mn}$ are the $n,m$ elements of the position operator, $E_k$ the energy of state $k$, and $N$ the number of electrons.  Note that the sum rules are consistent with Equation \ref{InvarientProduct}, i.e the product of the energy and square of the position operator is a constant if the number of electrons is fixed.  If electrons are added to the system in a way that does not change the energies, then the transition moment will be proportional to the square root of the number of electrons.  Specifically, Equation \ref{InvarientProduct} can be generalize to yield,
\begin{equation}\label{InvarientProductN}
\vec{r}\cdot \vec{r} E = k N,
\end{equation}
where $k$ is a constant.

Any component of the position operator, for example, $z = \hat{z} \cdot \vec{r}$, obeys the  sum rules given by Equation \ref{therule}.  By convention, we define the $\hat{x}$-direction to be along the largest diagonal component of the hyperpolarizability tensor.  Thus, by $\beta$ we implicitly mean $\beta_{xxx}$.  Furthermore, owing to Equation \ref{position}, we will loosely refer to $x_{nm}$ as the transition moment along the $\hat{x}$-direction between states $n$ and $m$.

Using the sum rules and the three-level ansatz, it can be shown that the polarizability, hyperpolarizability, and higher-order hyperpolarizability are bounded.\cite{kuzyk00.01,kuzyk00.02,kuzyk03.02,kuzyk03.01}  In particular, the fundamental limit of the off-resonant polarizability (also called the zeroth-order hyperpolarizability) is,
\begin{equation}\label{alphamax}
\alpha \leq \alpha_0^{max} = \left( \frac {e \hbar} {\sqrt{m}} \right)^2 \frac {N} {E_{10}^2};
\end{equation}
the fundamental limit of the hyperpolarizability (also called the first hyperpolarizability) is,
\begin{equation}\label{betamax}
\left|\beta_0 \right| \leq \beta_0^{max} =  \sqrt[4]{3} \left( \frac {e \hbar} {\sqrt{m}} \right)^3 \left[ \frac {N^{3/2}} {E_{10}^{7/2}} \right] ;
\end{equation}
and the fundamental limit of the second second hyperpolarizability, is,
\begin{equation}\label{gammamax}
-\left( \frac {e \hbar} {\sqrt{m}} \right)^4 \frac {N^2} {E_{10}^5} \leq \gamma_0 \leq 4 \left( \frac {e \hbar} {\sqrt{m}} \right)^4  \frac {N^2} {E_{10}^5} \equiv \gamma_0^{max}.
\end{equation}
Similarly, the transition moment to any excited state is bounded, and for the first excited state is given by,
\begin{equation}\label{x01MAX}
\left| x_{01} \right| \leq \sqrt{\frac {N \hbar^2} {2m E_{10}}} \equiv x_{01}^{MAX}.
\end{equation}
Clearly, the fundamental limit of the $n^{th}$ hyperpolarizability is of the form
\begin{equation}\label{hyperpolLimitForm}
\eta_{MAX}^{(n)} \propto \frac {N^{(n+2)/2}} {E_{10}^{(3n+4)/2} },
\end{equation}
where $\eta_{MAX}^{(-1)}$ is the fundamental limit of $x_{01}$, $\eta_{MAX}^{(0)}$ is the fundamental limit of $\alpha$, $\eta_{MAX}^{(1)}$ is the fundamental limit of $\beta$, etc.

Recall that the $n^{th}$ hyperpolarizability is of the form
\begin{equation}\label{nth-hyperpolarizabilty}
\eta^{(n)} = \frac {[x]^{n+2}} {[E]^{n+1}},
\end{equation}
where $[x]$ represents transition moments of the form $x_{nm}$ and $[E]$ represents an energy difference of the form $E_{m0} = E_m - E_0$.  Thus, re-scaling the $nth$ hyperpolarizability according to Equation \ref{InvarientProduct} and using Equation \ref{InvarientProductN} yields,
\begin{equation}\label{hyperpolTransform}
\eta^{(n)} = \frac {[x]^{n+2}} {[E]^{n+1}} =  \frac {k^{(n+2)/2}} {[E]^{(3n+4)/2}} N^{(n+2)/2}.
\end{equation}
The intrinsic $n^{th}$ hyperpolarizability is given by,
\begin{equation}\label{intrinsic}
\eta_{INT}^{(n)} = \frac {\eta^{(n)}} {\eta_{MAX}^{(n)}} \propto k^{(n+2)/2} \left( \frac {E_{10}} {\left[ E \right]} \right)^{(3n+4)/2},
\end{equation}
where we have used Equations \ref{hyperpolLimitForm} and \ref{hyperpolTransform}.  $\eta_{INT}^{(n)}$ is clearly unchanged under simple scaling and independent of the number of electrons.  Thus, $\eta_{INT}^{(n)}$ is a scale-invariant quantity.

The intrinsic hyperpolarizabilities are quantities that remove the effects of scaling, and can be used as a metric for comparing a molecule's nonlinear-optical efficiency, independent of the number of electrons or energy gap.  Simply stated, larger molecules with more electrons will generally interact more strongly with light than smaller, electron poor systems.  The intrinsic hyperpolarizabilities remove such effects, allowing one to focus on the structural properties that affect the response.  Only then can truly new paradigms be developed for making large molecules with exceptionally enhanced response.

\section{Using Scale-Invariance to Assess Molecules}\label{sec:application}

As described in the previous section, the increase of the hyperpolarizability of a molecule with size can be partially explained by the increase in the number of electrons and lowering of the energy gap (defined as the difference in energy between the ground and first excited state).  This dependence we shall refer to as {\em Simple Scaling}.  The intrinsic hyperpolarizability is a quantity that accounts for Simple Scaling, and can be used to compare molecules of all shapes and sizes.  A plot of the intrinsic hyperpolarizability as a function of a parameter of interest is referred to as a {\em Universal Plot}, and provides a bird's-eye view of the behavior of a large and diverse set of molecules.

\begin{figure}
\includegraphics{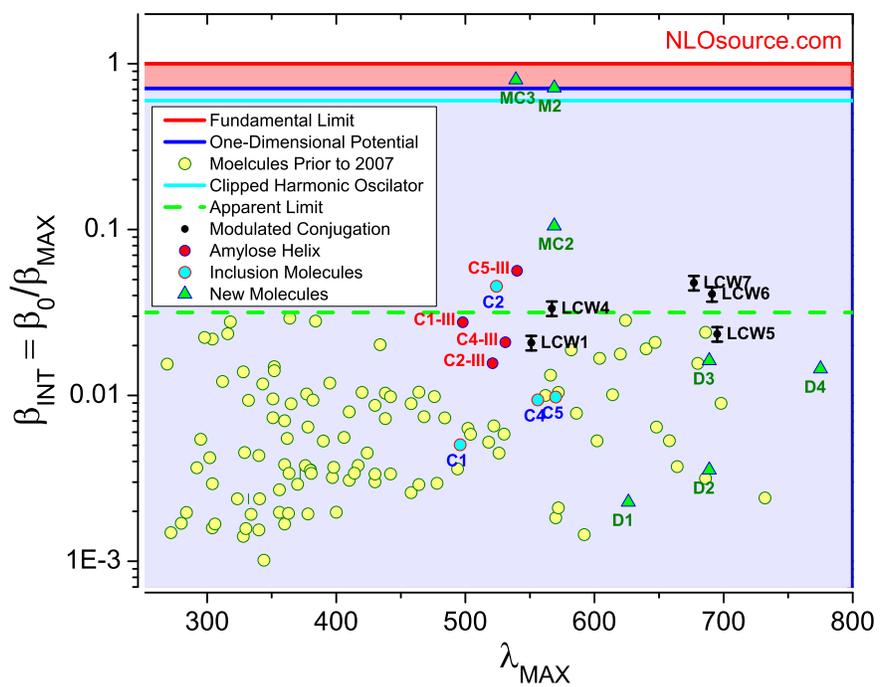}
\caption{A Universal Plot that includes all of the best measured molecules.  With permission from NLOsource.com}
\label{fig:UniversalPlot}
\end{figure}

Figure \ref{fig:UniversalPlot} shows a Universal Plot with $\lambda_{MAX}$ as the independent parameter.  The dashed green line, sometimes referred to as the {\em apparent limit}, represents an intrinsic hyperpolarizability of about 0.03.  They yellow points show the best molecules measured prior to 2007, all of which are below the apparent limit.  While three decades of intense research led to molecules with large hyperpolarizabilities, their intrinsic hyperpolarizabilities showed little improvement.  Thus, all of the research efforts prior to about 2007 that were  aimed at improving the nonlinear-optical response essentially lead to larger electron-rich molecules with smaller energy gap.

Molecules D1 through D4 have large hyperpolarizabilities, and are used in making electrooptic devices.\cite{shi00.01}  However, the Universal Plot clearly shows that the intrinsic hyperpolarizabilities are unremarkable.  Indeed, molecule D4 has a large hyperpolarizability owing mostly to its small energy gap ($\lambda_{MAX}$ is the largest of the group) and large number of electrons.  If it were possible to adjust the structure of such molecules to attain an intrinsic hyperpolarizability of unity, the hyperpolarizability could be improved one hundred fold.
\begin{figure}
\includegraphics{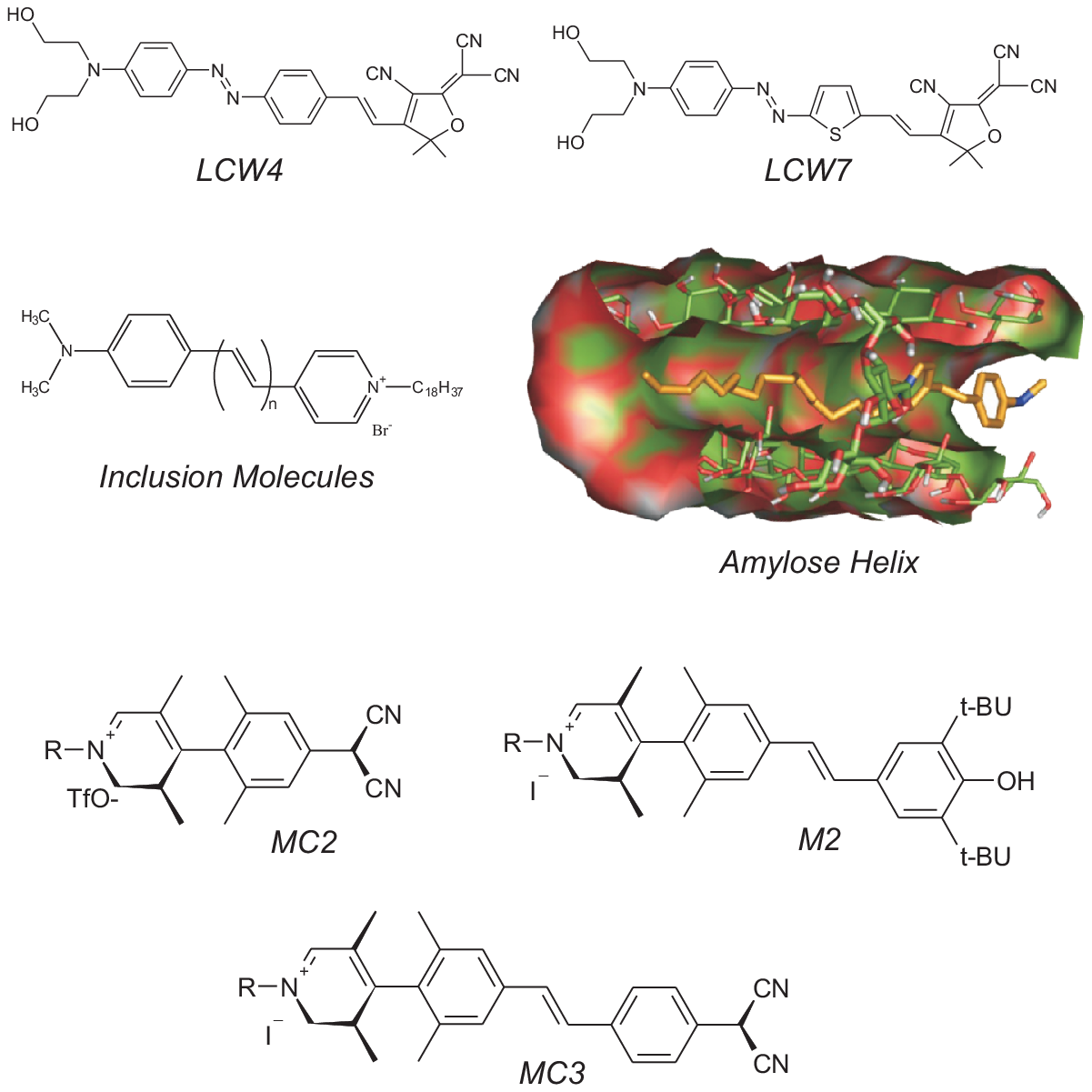}
\caption{Some representative molecules from the Universal Plot shown in Figure \ref{fig:UniversalPlot}.}
\label{fig:Molecules}
\end{figure}

In an effort to design new paradigms for breaking through the apparent limit, Zhou et al used numerical optimization techniques to determine what potential energy functions would lead to the largest hyperpolarizability.\cite{zhou06.01,zhou07.02} They found that there were many one-dimensional potential energy functions that lead to intrinsic hyperpolarizabilities of 0.709, but not one was found to be larger.  In all cases, the three-level ansatz was obeyed.

One class of potentials that optimize the hyperpolarizability have random (i.e. non-periodic) undulations that appear to serve the purpose of spatially separating the wavefunctions to insure that the three-level ansatz in obeyed.\cite{zhou07.02} As such, Zhou and coworkers proposed that modulation of conjugation might be a new paradigm for making better molecules.  The most common molecules with large hyperpolarizability where electron donors and acceptors separated by a conjugated bridge of the form C--C=C--C-C=C--C=C.  The concept of modulated conjugation suggests that bridges with differing atoms, such as C=N--C=S--C... etc., might yield a larger response.

P\'{e}rez-Moreno and coworkers showed that molecules with a modulated bridge, as shown by the points labeled LCW6 and LCW7 in Figure \ref{fig:UniversalPlot}, had larger intrinsic hyperpolarizability than the non-modulated versions, labeled LCW4.\cite{perez07.01,perez09.01} While it is difficult to theoretically associate an effective potential energy profile of these molecules (shown in Figure \ref{fig:Molecules}), they argue that differences in aromatic stabilization energy of the rings serve as a proxy for modulated conjugation.

As the molecules in the series labeled C1 to C5 are made longer (see Figure \ref{fig:Molecules}), the intrinsic hyperpolarizability at first increases, then decreases as shown in Figure \ref{fig:UniversalPlot}.\cite{perez07.02}  The decrease is attributed to the inability of the larger molecules to retain planarity.  To test this hypothesis, each molecule in the series was placed in an amylose helix, which keeps the molecules planar.  The measured hyperpolarizabilities are labeled C1-III to C5-III.  The longest molecule, C5-III, when place in the amylose helix, has the largest intrinsic  hyperpolarizability.  Its absolute hyperpolarizability is enhanced by about a factor of ten over the non-encapsulated form.

Kang and Coworkers considered twisted molecules, as shown in the bottom of Figure \ref{fig:Molecules}.\cite{Kang05.01,Kang07.01} The points in Figure \ref{fig:UniversalPlot} labeled M2, MC2, and MC3 clearly have the largest intrinsic hyperpolarizabilities ever measured.  Thus, this paradigm shows great promise in making molecules with exceptional hyperpolarizability if the enhancement due to the twist can be maintained for larger forms of such a molecule.

\section{New Paradigms for Making Molecules}\label{paradigms}

Aside from the twisted molecules, the hyperpolarizabilities of most molecules fall far short of the fundamental limit.  Even some numerical optimization methods seem to be bound by the apparent limit.  For example, Keinan, Wang and coworkers use a computer algorithm to sort through a library containing about a million building blocks of linear combinations of atomic potentials to find structures with large nonlinear-optical response.\cite{wangm06.01,keina08.01}  They have found calculated off-resonant hyperpolarizabilities on the order of $\beta_0 = 10,000 \times 10^{-30} esu$; but, the intrinsic hyperpolarizability is $\beta_0^{int} = 0.013$, more than a factor of 2 from the apparent limit and just over 1\% of the fundamental limit.  If the intrinsic hyperpolarizability were optimized so that all of the electrons were used efficiently, the hyperpolarizability would be about $\beta_0 = 800,000 \times 10^{-30}\,esu$.

In contrast, our approach is to use abstract numerical optimization, which apply to any quantum system, with the goal of identifying common features that are shared by systems whose nonlinear response is at the fundamental limit.  While this approach may not lead to specific molecular structures, it can provide general guidelines.  Our numerical optimization studies have found certain universal properties of a quantum system when the intrinsic hyperpolarizability $\beta_{xxx}^{INT}$ is optimized.  They are as follows:

When the intrinsic hyperpolarizability $\beta_{xxx}^{INT}$  is optimized by varying the potential energy function or the geometry of point charges in 2-dimensions,
\begin{itemize}
\item{the ratio of energies is always about $E_{20}/E_{10} = 0.49$;}
\item{the transition moment to the dominant state is given by $x_{10} / x_{10}^{MAX} = 0.79$;}
\item{the optimized intrinsic hyperpolarizability is always $0.709$ for a broad range of potentials;}
\item{exactly three states dominate the nonlinear-optical response;}
\item{and, the electron cloud of the three dominant states tends to be aligned in one dimension.}
\end{itemize}

These observation give clues about how to design better molecules, and introduce new puzzles.  For example, searches for better materials should seek molecules with approximately equally-spaced energy levels and two dominant excited states.  One design paradigm, suggested above, is the synthesis of new molecules that exhibit modulation of conjugation.  New theoretical studies, on the other hand, should focus on identifying ways of arranging atoms in a molecule in such as way as to provide an effective potential along the conjugated pathway with the kinds of undulations that cause the eigenfunctions to be spatially separated in a way that enforces the three-level ansatz.

\begin{figure}
\includegraphics{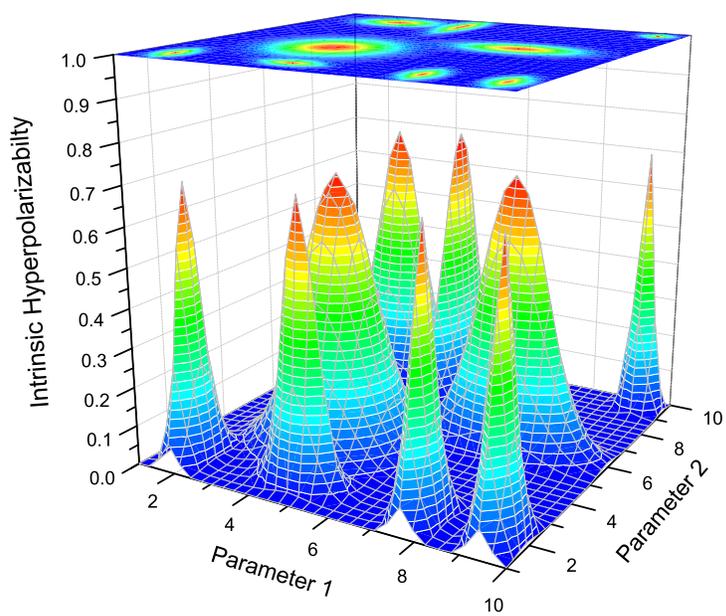}
\caption{The intrinsic hyperpolarizability landscape, which abstractly represents the fact that many optimized potential energy functions lead to the same intrinsic hyperpolarizability.}
\label{fig:Landscape}
\end{figure}

While these universally-observed properties might be used to design better molecules, the fact that universal properties exist is puzzling.  Why are there so many different potentials that all yield the same optimized $\beta_{xxx}^{INT}$?  Why is the optimized value given by $\beta_{xxx}^{INT} = 0.709$, and not unity?  Figure \ref{fig:Landscape} shows a two-dimensional representation of the hyperpolarizability landscape, where Parameter 1 and Parameter 2 quantify the potential energy function used to calculate the hyperpolarizability.  The same kind of landscape is found for other quantities, such as the energy ratio.  Why do such a diverse range of potentials all yield the same energy ratio and normalized transition moment?  Answers to such questions may lead to a better understanding of light-matter interaction; but, may also lead to new fundamental principles.

\section{Conclusion}

We have shown that the fundamental limit of the $n^{th}$ hyperpolarizability defines an absolute standard for comparing molecules.  The ratio of a quantum system's $n^{th}$ hyperpolarizability to the fundamental limit defines the intrinsic $n^{th}$ hyperpolarizability, which is a scale-invariant quantity that can be used to generate a Universal Plot of the nonlinear response of a broad range of molecules.  Since this Universal Plot takes into account Simple Scaling, it can be used to better focus on the relevant properties required to develop new paradigms for understanding structural properties that enhance the response.

All quantum systems with optimized intrinsic hyperpolarizabilities are dominated by two excited states; and, no system has been observed with an intrinsic hyperpolarizability near unity with more than 2 dominant excited states.  An investigation of the potential energy functions that optimizes $\beta_{xxx}$ has led to the proposal that systems with conjugated modulation may be a new paradigm for large-hyperpolarizability molecules.  Limited experimental evidence supports this hypothesis.  Other paradigms that appear to enhance the hyperpolarizability are twisted molecules; and, in the other extreme, using an amylose helix inclusion complex to force large chromophores to be planar.

For one-dimensional quantum systems that are representable with a potential energy function, the ratio of the first to second excited state energy is found to be about 0.49 and the normalized transition moment to the first excited state is 0.79.  Thus, it may be possible to identify candidate molecules by screening their linear optical properties for these values.  More importantly, these results hint of a set of universal properties that are required of optimized molecules.  The fact that many different systems all share these universal properties is a puzzle whose solution may lead to a deeper understand of the nonlinear-optical response of a quantum system.

{\bf Acknowledgements}: MGK thanks the National Science Foundation (ECCS-0756936) and Wright Paterson Air Force Base for generously supporting this work.

\section{Appendix}

In this appendix, we show that the commutator given by Equation \ref{commutator} follows from the general Hamiltonian given by Equation \ref{generalHamiltonian}.  We do so by considering each term in the Hamiltonian.

The potential $V \left(\vec{r}_1, \dots \vec{r}_N; \vec{s}_1, \dots \vec{s}_N ; \vec{L}_1, \dots \vec{L}_N\right)$. i.e. the last term in Equation \ref{generalHamiltonian}, can include many types of interactions.  For potentials of the form $V \left(\vec{r}_1, \dots \vec{r}_N; \vec{s}_1, \dots \vec{s}_N \right)$, $[x,V]=0$.  Spin-orbit coupling contributions have terms of the form $v \vec{S}\cdot \vec{L}$, where it is understood that $v$ is a function of the electron coordinates.  The commutator with $x$ yields,
\begin{equation}\label{LastTerm}
\left[x, v \vec{S}\cdot \vec{L} \right] =  v \vec{S} \cdot \left[x, \vec{L} \right] + v \left[x, \vec{S} \right] \cdot \vec{L} = f(\vec{r}),
\end{equation}
where $f(\vec{r})$ is a function of position.  We have used the fact that $\left[x, \vec{S} \right] = 0$ because $\vec{S}$ and $\vec{r}$ are operators in orthogonal sub-spaces; and that $\left[x, \vec{L} \right]$ has terms of the form
\begin{equation}\label{LastTerm2}
\left[x, L_z \right] = \left[x, x p_y - y p_x \right] = -\left[x, y p_x \right] = -y \left[x, p_x \right] = -y i \hbar.
\end{equation}
Clearly, Equation \ref{LastTerm2} gives terms that are only functions of the position.  For terms of this type,
\begin{equation}\label{firstAll}
\left[x,\left[ x,V \right] \right] = 0.
\end{equation}
Thus, Equation \ref{firstAll} generally holds.  However, we note that it is possible to pick potential functions that will yield $\left[x,\left[ x,V \right] \right] \neq 0$, but none of these are physically meaningful.

Next we consider the commutator of first term in Equation \ref{generalHamiltonian} with x.  Expanding the square yields three terms,
\begin{equation}\label{FirstTerm1}
T_1 = \left[x, p^2  \right] = 2i \hbar p_x,
\end{equation}
\begin{equation}\label{FirstTerm2}
T_2 = \left[x, \vec{p} \cdot \vec{A} +  \vec{A} \cdot \vec{p} \right],
\end{equation}
and
\begin{equation}\label{FirstTerm3}
T_3 = \left[x, A_2 \right] = 0.
\end{equation}
Each term in Equation \ref{FirstTerm2} is a function of the vector potential, as follows,
\begin{equation}\label{FirstTerm2com}
\left[x, \vec{p} \cdot \vec{A} \right] = \vec{p} \cdot \left[x, \vec{A} \right] + \left[x, \vec{p} \right] \cdot \vec{A} = 0 + i \hbar A_x.
\end{equation}

Using Equations \ref{FirstTerm1} through \ref{FirstTerm2com} yields
\begin{equation}\label{FirstTermTogether}
\left[x,\left[x , \frac {1} {2m} \sum_k^{N} \left( \vec{p}_k - \frac {e} {c} \vec{A} \left(\vec{r}_1, \dots \vec{r}_N \right) \right)^2 \right]\right] = \frac {\hbar^2} {m}.
\end{equation}
Equations \ref{firstAll} and \ref{FirstTermTogether} lead to Equation \ref{commutator}.


\begin{thebibliography}{10}
\newcommand{\enquote}[1]{``#1''}
\expandafter\ifx\csname url\endcsname\relax
  \def\url#1{{#1}}\fi
\expandafter\ifx\csname urlprefix\endcsname\relax\def\urlprefix{}\fi

\bibitem{kuzyk00.01}
M.~G. Kuzyk, \enquote{{Physical Limits on Electronic Nonlinear Molecular
  Susceptibilities},} Phys. Rev. Lett. {\bf 85}, 1218 (2000).

\bibitem{kuzyk00.02}
M.~G. Kuzyk, \enquote{{Fundamental limits on third-order molecular
  susceptibilities},} Opt. Lett. {\bf 25}, 1183 (2000).

\bibitem{orr71.01}
B.~J. Orr and J.~F. Ward, \enquote{{Perturbation Theory of the Non-Linear
  Optical Polarization of an Isolated System},} Molecular Physics {\bf 20},
  513--526 (1971).

\bibitem{kuzyk05.01}
M.~G. Kuzyk, \enquote{{Reply to comment on "Physical Limits on Electronic
  Nonlinear Molecular Susceptibilities"},} Phys. Rev. Lett. {\bf 95}, 109\,402
  (2005).

\bibitem{zhou06.01}
J.~Zhou, M.~G. Kuzyk, and D.~S. Watkins, \enquote{{Pushing the
  hyperpolarizability to the limit},} Opt. Lett. {\bf 31}, 2891 (2006).

\bibitem{kuzyk06.03}
M.~G. Kuzyk, \enquote{{Fundamental limits of all nonlinear-optical phenomena
  that are representable by a second-order susceptibility},} J. Chem Phys. {\bf
  125}, 154\,108 (2006).

\bibitem{zhou07.02}
J.~Zhou, U.~B. Szafruga, D.~S. Watkins, and M.~G. Kuzyk, \enquote{{Optimizing
  potential energy functions for maximal intrinsic hyperpolarizability},} Phys.
  Rev. A {\bf 76}, 053\,831 (2007).

\bibitem{kuzyk08.01}
M.~C. Kuzyk and M.~G. Kuzyk, \enquote{{Monte Carlo Studies of the Fundamental
  Limits of the Intrinsic Hyperpolarizability},} J. Opt. Soc. Am. B. {\bf 25},
  103--110 (2008).

\bibitem{zhou08.01}
J.~Zhou and M.~G. Kuzyk, \enquote{{Intrinsic Hyperpolarizabilities as a Figure
  of Merit for Electro-optic Molecules},} J. Phys. Chem. C. {\bf 112},
  7978--7982 (2008).

\bibitem{kuzyk01.01}
M.~G. Kuzyk, \enquote{{Quantum limits of the hyper-Rayleigh scattering
  susceptibility},} IEEE Journal on Selected Topics in Quantum Electronics {\bf
  7}, 774 --780 (2001).

\bibitem{kuzyk03.02}
M.~G. Kuzyk, \enquote{{Erratum: Physical Limits on Electronic Nonlinear
  Molecular Susceptibilities},} Phys. Rev. Lett. {\bf 90}, 039\,902 (2003).

\bibitem{kuzyk03.01}
M.~G. Kuzyk, \enquote{{Fundamental limits on third-order molecular
  susceptibilities: erratum},} Opt. Lett. {\bf 28}, 135 (2003).

\bibitem{shi00.01}
Y.~Shi, C.~Zhang, H.~Zhang, J.~H. Bechtel, L.~R. Dalton, B.~H. Robinson, and
  W.~H. Steier, \enquote{{Low (Sub-1-Volt) Halfwave Voltage Polymeric
  Electro-optic Modulators Achieved by Controlling Chromophore Shape},} Science
  {\bf 288}, 119--122 (2000).

\bibitem{perez07.01}
J.~P\'{e}rez-Moreno, Y.~Zhao, K.~Clays, and M.~G. Kuzyk, \enquote{{Modulated
  conjugation as a means for attaining a record high intrinsic
  hyperpolarizability},} Opt. Lett. {\bf 32}, 59--61 (2007).

\bibitem{perez09.01}
J.~P\'{e}rez-Moreno, Y.~Zhao, K.~Clays, M.~G. Kuzyk, Y.~Shen, L.~Qiu, J.~Hao,
  and K.~Guo, \enquote{Modulated Conjugation as a Means of Improving the
  Intrinsic Hyperpolarizability,} J. Am. Chem. Soc. {\bf 131}, 5084–5093
  (2009).

\bibitem{perez07.02}
J.~P\'{e}rez-Moreno, I.~Asselberghs, Y.~Zhao, K.~Song, H.~Nakanishi, S.~Okada,
  K.~Nogi, O.-K. Kim, J.~Je, J.~Matrai, M.~De~Mayer, and M.~G. Kuzyk,
  \enquote{{Combined molecular and supramolecular bottom-up nano-engineering
  for enhanced nonlinear optical response: Experiments, modelling and
  approaching the fundamental limit},} J. Chem. Phys. {\bf 126}, 074\,705
  (2007).

\bibitem{Kang05.01}
H.~Kang, A.~Facchetti, P.~Zhu, H.~Jiang, Y.~Yang, E.~Cariati, S.~Righetto,
  R.~Ugo, C.~Zuccaccia, A.~Macchioni, C.~L. Stern, Z.~Liu, S.~T. Ho, and T.~J.
  Marks, \enquote{{Exceptional Molecular Hyperpolarizabilities in Twisted
  $\pi$-Electron System Chromophores},} Angew. Chem. Int. Ed. {\bf 44} (2005).

\bibitem{Kang07.01}
H.~Kang, A.~Facchetti, H.~Jiang, E.~Cariati, S.~Righetto, R.~Ugo, C.~Zuccaccia,
  A.~Macchioni, C.~L. Stern, Z.~F. Liu, S.~T. Ho, E.~C. Brown, M.~A. Ratner,
  and T.~J. Marks, \enquote{{Ultralarge hyperpolarizability twisted
  $\pi$-electron system electro-optic chromophores: Synthesis, solid-state and
  solution-phase structural characteristics, electronic structures, linear and
  nonlinear optical properties, and computational studies},} J. Am. Chem. Soc.
  {\bf 129}, 3267--3286 (2007).

\bibitem{wangm06.01}
M.~Wang, X.~Hu, D.~N. Beratan, and W.~Yang, \enquote{{Designing molecules by
  optimizing potentials},} J. Am. Chem. Soc. {\bf 128}, 3228--3232 (2006).

\bibitem{keina08.01}
S.~Keinan, M.~J. Therien, D.~N. Beratan, and W.~T. Yang, \enquote{{Molecular
  Design of Porphyrin-Based Nonlinear Optical Materials},} J. Phys. Chem. A
  {\bf 112}, 12\,203--12\,207 (2008).

\end{thebibliography}

\end{document}